\documentclass{PoS}
\usepackage[utf8]{inputenc}

\usepackage[table,x11names]{xcolor}
\usepackage{amssymb}	
\usepackage{booktabs,amsmath} 
\usepackage{bbm}
\usepackage{subcaption}
\usepackage{amsfonts}
\usepackage[skip=0.5\baselineskip]{caption}

\usepackage{geometry} 
\definecolor{boxcolor}{HTML}{ffe6a1}

\def \Ns {{N_{\sigma}}}
\def \Nt {{N_{\tau}}}
\def \Nc {{N_{c}}}

\newcommand{\lr}[1]{\left( #1 \right)}

\let\OLDthebibliography\thebibliography
\renewcommand\thebibliography[1]{
  \OLDthebibliography{#1}
  \setlength{\parskip}{0pt}
  \setlength{\itemsep}{0pt plus 0.3ex}
}

\title{Continuous Time Simulations of Strong Coupling LQCD at Finite Baryon Density}

\ShortTitle{CT Simulations of SCLQCD at finite $\mu_B$}

\author{\speaker{Marc Klegrewe}\\
Bielefeld University\\
        E-mail:  \email{mklegrewe@physik.uni-bielefeld.de}}
\author{Wolfgang Unger\\
Bielefeld University\\
E-mail: \email{wunger@physik.uni-bielefeld.de}}

\abstract{We study lattice QCD in the limit of infinite gauge coupling on a discrete spatial yet continuous Euclidean time lattice at finite baryon chemical potential $\mu_B$. The continuous time framework is based on sending $N_\tau\rightarrow \infty$ and the bare anisotropy to infinity while fixing the temperature in a non-perturbative setup. This leads to a sign problem free algorithm that allows us to study the whole $\mu_B$-$T$ plane. We construct Taylor coefficients required for a Taylor expansion in the pressure at zero chemical potential and in the chiral limit. On that account, cumulants in the baryon number density are measured in various fashions to improve on accuracy. These calculations are based on our worm type Monte Carlo algorithm featuring a polymer resummation scheme and a histogram method.}

\FullConference{The 37th Annual International Symposium on Lattice Field Theory - LATTICE2019\\
		16-22 June, 2019\\
		Central China Normal University, Wuhan, Hubei, China.}

\begin{document}
\section{Motivation}
The QCD phase diagram illustrates the states of strongly coupled matter under extreme conditions. Lattice QCD is the tool of choice within this non-perturbative setup where the transition region separating hadronic matter from the quark-gluon plasma (QGP) is of vast interest. Unfortunately, exploring the rich phase diagram structure is hindered by the severe complex phase problem which arises at finite baryon chemical potentials $\mu_B$ since the fermion determinant becomes complex and importance sampling is no longer applicable. There is a large pool of methods available to circumvent the sign problem. As for the Taylor expansion method, calculations are performed at zero chemical potential and results expanded to finite chemical potential \cite{Allton2005}. Alternatively, one can exploit the fact that the sign problem is representation dependent. Hence, dual representations are of interest where the sign problem is milder or vanishes completely. In the infinite coupling limit ($g\rightarrow \infty; \beta=\frac{2N_c}{g^2}\rightarrow 0$) such a representation is known. This approach is based on integrating out the gauge degrees of freedom exactly, resulting in color singlets. Finally - after Grassmann integration - the partition function is expressed as a gas of hadron world lines (c.f. monomer-dimer system \cite{rossi1984}).

Within this particular framework, the complete phase diagram can be investigated. It shares important features of QCD at finite $\beta$ such as confinement and spontaneous chiral symmetry breaking and its restoration at a transition temperature $T_c$. Within this scheme, simulations in the chiral limit are very economical and faster compared to Hybrid Monte Carlo due to the applicability of the worm algorithm \cite{forcrand2010}. Further assets are acquired when formulating the Continuous Euclidean Time (CT) limit framework within the strong coupling limit of LQCD (SC-LQCD) \cite{Klegrewe2018}. We benefit from a complete absent sign problem and observables that have a particular simplified structure. 

We use these features to construct a Taylor expansion in the described framework based on a histogram method which gives access to the calculation of Taylor coefficients to high orders. We profit from the intrinsic feature of the CT limit of having static baryons combined with a polymer resummation scheme which also incorporates static meson lines. We can probe all results for accuracy, as Monte Carlo simulations at finite density are unrestricted.
\section{Strong Coupling QCD on anisotropic lattices}
At infinite gauge coupling $g\rightarrow\infty$ the coefficient of the plaquette term $\beta=6/g^2$ vanishes. Thus, the Yang Mills part $F_{\mu\nu} F_{\mu\nu}$ is absent which allows to analytically integrate out the gauge fields in the Dirac operator. In the SC limit no lattice spacing can be defined as it becomes maximally coarse and the continuum limit is by far out of reach. After integration over the Grassmann fields the SC partition function for staggered fermions and for a discrete system confined in a volume of size $\Ns^3\times \Nt$ is given by
\begin{align}
Z(\gamma,N_\tau,m_q,\mu_q)=& \sum_{\{k,n,\ell\}}\prod_{x,\mu}\frac{(\Nc-k_{\mu}(x))!}{\Nc!k_{\mu}(x)!}\gamma^{2k_{\mu}(x)\delta_{0,\mu}}\prod_{x}\frac{\Nc!}{n_x!}(2am_q)^{n_x} \prod_\ell w(\ell,\mu_q) \label{SCPF}\\
\text{Grassmann constraint:} & \quad n_x+\sum_{\mu=\pm 0,\ldots \pm d} \left( k_{\mu}(x) + \frac{N_c}{2} |\ell_\mu(x)| \right)=\Nc, \quad 
\forall x\in \Ns^3\times \Nt \label{GC} \\
w(\ell,\mu_q)=\sigma(\ell)\gamma^{\Nc \sum_x|\ell_0(x)|} & \exp\lr{\Nc \Nt r(\ell) a_\tau \mu_q}, \qquad \sigma(\ell)=(-1)^{r(\ell)+N_-(\ell)+1}\prod_{b=(x,\mu)\in \ell}\eta_\mu(x).\label{loops}
\end{align}
Note that the SC partition function follows exactly from the above described procedure of interchanging the integration order. We have confined, colorless, discrete dual degrees of freedom:
\begin{itemize}
\setlength\itemsep{0em}
\item Mesonic degrees of freedom $k_{\mu}(x)\in \{0,\ldots,N_c\}$ (non-oriented meson hoppings called dimers) and
$n(x) \in \{0,\ldots N_c\}$ (mesonic sites called monomers).
\item Baryonic degrees of freedom, which form oriented baryon loops $\ell$ with sign $\sigma(\ell)=\pm 1$ and winding number $r(\ell)$ that depend on the geometry of the loops  Eq.~(\ref{loops}).
These loops are self-avoiding and do not touch the mesonic degrees of freedom.
\end{itemize}
Both mesonic and baryonic degrees of freedom obey the Grassmann constraint Eq.~(\ref{GC}). As we restrict to the chiral limit $m_q=0$, the mesonic sector is purely described by dimers. 

The bare anisotropy coupling $\gamma$ in Eq.~(\ref{SCPF}) allows calculations on anistropic lattices with $\xi=\frac{a_\sigma}{a_\tau}$. The temperature is then defined as follows: 
\begin{align}
T=\frac{1}{a_\tau \tau}=\frac{\xi(\gamma)}{a N_\tau}.
\end{align}
If we now proceed towards the CT formulation, we find that the anisotropy parameter in terms of $\gamma$ reads $\xi(\gamma)\equiv \kappa\gamma^2$ with $\kappa$ being a proportionality constant which can be measured in CT. Finally, $\gamma$ and $N_\tau$ can be eliminated completely by the temperature which is encoded in the definition of the \emph{Continuous Euclidean Time Limit}:
\begin{align}
\Nt\rightarrow \infty, \qquad \gamma \rightarrow \infty, \qquad \kappa\gamma^2/\Nt\equiv aT \;\; {\rm fixed.}
\label{Eq:defCT}
\end{align}
Only one parameter is left that sets the thermal properties, and all discretization errors introduced by a finite $N_\tau$ are removed.
\section{Continuous Time Limit}
Simulations in the CT limit have the advantage that the temperature $aT$ is the only parameter: There is no need of extrapolations in the temporal lattice extent $\Nt \rightarrow \infty$ and the functional dependence of observables simplifies significantly since no anisotropy has to be accounted for. Both mesonic and baryonic degrees of freedom experience some changes: Temporal dimers are replaced by meson occupation numbers. Additionally, the CT limit enforces baryons to be static for $N_c\geq 3$: baryonic lines loop around the lattice in temporal direction without any spatial hoppings occuring. The sign problem is completely absent as seen in Eq.~(\ref{loops}) with $N_-(\ell)$ even valued and $r(\ell)=1$. Note that the baryon mass remains finite as baryons can be considered as non-relativistic particles.

It is possible to formulate the CT partition function in terms of a quantum Hamiltonian \cite{Klegrewe2019}
\begin{align}
\begin{split}
Z_{CT}(T,\mu_B)=\text{Tr}_{h}\left[ e^{(\hat{H}+\hat{\Omega}_B\mu_B)/T} \right], \quad \hat{H}=\frac{1}{2}\sum_{\langle \vec{x},\vec{y} \rangle} \left( \hat{J}^+_{\vec{x}}\hat{J}^-_{\vec{y}}+\hat{J}^-_{\vec{x}}\hat{J}^+_{\vec{y}} \right), \quad \hat{\Omega}_B=\sum_{\vec{x}}\hat{\omega}_x
\label{ZCT}
\end{split}
\end{align}
with $\hat{J}^{+/-}$ being the meson state raising and lowering operators and thus, $\hat{H}$ corresponds to the meson exchange. Furthermore, $\hat{\omega}_x$ is the baryon winding number at site $x$ and $\Omega_B$ is the baryon number operator. The trace runs over the hadronic quantum states $h$. $Z_{CT}$ can be expanded in a diagrammatic fashion into meson hoppings where the number of interactions at a given temperature is associated to the number of spatial dimer \begin{align}
N_{D_s}=\sum_x\sum_i k_i(x).
\end{align}
A deeper analysis on the corresponding algebra of the mesonic operators as well as the $N_f=2$ formulation is soon to be published in \cite{Klegrewe2019}. To sample this partition function a worm type algorithm is used, similar to the directed path algorithm introduced for SC-QCD in \cite{adams2003}.

As addressed before observables simplify greatly in the CT limit. We want to measure pressure and energy density which are related to the spatial dimer density $n_{D_s}$ as
\begin{align}
\begin{split}
a^4 p = -\frac{1}{3}\langle n_{D_s} \rangle, \quad a^4 \epsilon = -\langle n_{D_s} \rangle, \quad n_{D_s}=\frac{N_{D_s}}{N_\sigma^3}.
\end{split}
\end{align}
In the chiral limit of SC-LQCD the interaction measure is trivial: $\epsilon-3p=0$. Thus, $\epsilon$ and $p$ are proportional.
Furthermore, the Taylor expansion will require measurements of the baryon number density given as
\begin{align}
a^3n_B=a^3\frac{T}{V}\frac{\partial \ln \mathcal{Z}}{\partial \mu_B}\biggr\rvert_{V,T}=\langle \omega \rangle, \quad \omega=\frac{\Omega_B}{N_\sigma^3}=\frac{1}{N_\sigma^3}\sum_\ell r(\ell),
\label{Eq:baryonNumberDensity}
\end{align}
with the total winding number per configuration $\Omega_B=$($\#$baryons-$\#$anti-baryons)$=B-A$ and its density $\omega$, where $r(\ell)$ in Eq.~(\ref{loops}) simplifies to $\pm 1$ in the CT limit. Note that the volume $V$ in CT is implicitly given by $V=a^3N_\sigma^3$.
\section{Taylor expansion of the pressure}
In the following, we will probe the Taylor expansion method with finite density Monte Carlo results obtained by our CT worm algorithm. As an observable we study the pressure which is natively defined as
\begin{align}
a^4p=a^3 T\frac{\partial \ln Z}{\partial V}.
\label{Eq:nativePressure}
\end{align}
However, we assume the system to be homogeneous such that the volume derivative becomes merely a factor $V$. Finally, the expansion about $(T,\mu_B=0)$ follows as
\begin{align}
\label{Eq:taylorExpansionPressure}
\begin{split}
a^4 p=a^3\frac{T}{V}\ln Z=a^4 p(T,\mu_B=0) + \sum_{n=1}^{\infty} c_{2n}\left( \frac{\mu_B}{T} \right)^{2n}, \quad \text{with} \quad c_{2n}=a^3\frac{T}{V}\frac{1}{(2n)!}\frac{\partial^{2n}\ln Z}{\partial(\mu_B/T)^{2n}}\biggr\rvert_{\mu_B=0}.
\end{split}
\end{align}
The pressure is only an even function of $\mu_B$ due to the reflection symmetry $P(T,\mu_B)=P(T,-\mu_B)$. The essential contribution comes from $\frac{\partial^{2n}\ln Z}{\partial(\mu_B/T)^{2n}}$. Since the partition function $Z$ is the moment generating function, and hence, its logarithm $\ln Z$ is the cumulant generating function, we find the relation
\begin{align}
\frac{\partial^{2n}\ln Z}{\partial(\mu_B/T)^{2n}}=\kappa_{2n}(\Omega_B)=\kappa_{2n}(\omega) \cdot V_\sigma^{2n}, \quad V_\sigma=N_\sigma^3.
\label{Eq:logZderivativeKappa}
\end{align}
We introduced the cumulants $\kappa_{2n}$ in the winding number or its density. The latter can be constructed from measurements of moments of baryon number densities as implied by Eq.~(\ref{Eq:baryonNumberDensity}). The cumulants are expressed in terms of moments as
\begin{align}
\begin{split}
\kappa_1(\omega) & = \mu_1, \\
\kappa_2(\omega) & = \mu_2-\mu_1^2, \\
\kappa_3(\omega) & = \mu_3-3\mu_2\mu_1+2\mu_1^3, \\
\kappa_4(\omega) & = -6\mu_1^4+12\mu_1^2\mu_2-3\mu_2^2-4\mu_1\mu_3+\mu_4, \\
\dots
\label{Eq.kappaDefinition}
\end{split}
\end{align}
with $\mu_m$ related to be baryon number densities via $\mu_m=\langle \left(\frac{B-A}{V_\sigma}\right)^m\rangle=\langle \omega^m\rangle$. At zero baryon chemical potential $\mu_B=0$ there is no favoring between the baryon and anti-baryon sector resulting in odd baryon density moments to vanish. Thus, the expressions in Eq.~(\ref{Eq.kappaDefinition}) are simplified.

Putting all the insights together reduces the expression for the Taylor expansion of the pressure to
\begin{align}
\label{Eq:defineTayloCoeff}
\begin{split}
a^4\Delta p = a^4\left(p(T,\mu_B)-p(T,0)\right) =  \sum_{n=1}^{\infty}c_{2n}(\Omega_B)\left(\frac{\mu_B}{T}\right)^{2n} =a^3 \frac{T}{V}\sum_{n=1}^{\infty}\frac{1}{(2n)!}\left(\frac{\mu_B}{T}\right)^{2n} \kappa_{2n}(\Omega_B).
\end{split}
\end{align}
Within the next section we summarize our results on measuring $a^3\kappa_{2n}(\Omega_B)/V_\sigma$ to the highest possible order. The division by volume is necessary in order to match the high-temperature limit.
\section{Measurements of Cumulants}
\begin{figure}
\centering
\includegraphics[width=0.80\textwidth]{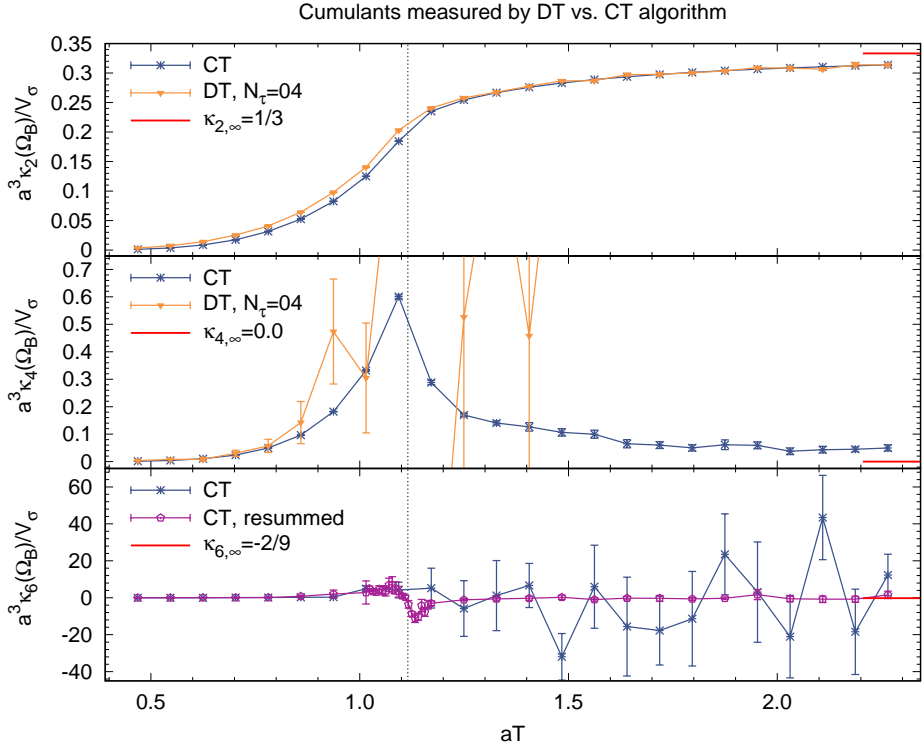}
\caption{Measurement of the cumulants $\kappa_2$, $\kappa_4$ and $\kappa_6$ on a $10^3\times CT$ lattice. Discrete time simulations fail to yield reasonable results for $\kappa_4$ while the standard CT algorithm starts to struggle with $\kappa_6$. A polymer resummation brings all cumulants up to $\kappa_6$ under control but not beyond.}
\label{Fig:kappa2_4_6}
\end{figure}

The  naive calculation of Taylor coefficients is analogous to measuring moments in the baryon number density $n_B$. This is not CT specific and was already discussed in \cite{Bollweg2018} for discrete time SC-LQCD. Within this framework it already fails to extract reasonable results for $\kappa_4$ (Fig.~\ref{Fig:kappa2_4_6}). In contrast, CT data strive from the inherent algorithm feature of having static baryons.
Nevertheless, the usual CT algorithm breaks down at the rather low order $\kappa_6$ even if the polymer resummation is included: In the chiral limit, we can resum baryons and anti-baryons ($P$-Polymers), and additionally include static meson lines ($Q$-Polymers). The corresponding weights are
\begin{align}
w_P(\mu_B/T)=2 \cosh(\mu_B/T),\quad w_Q(\mu_B/T) =N_c+1+2 \cosh(\mu_B/T).
\label{PolymerWeight}
\end{align}
This procedure leads to a significant improvement for $\kappa_6$ (Fig.~\ref{Fig:kappa2_4_6}). Even though it is possible to extract a reasonable signal with more statistics, we would still fail to address higher order cumulants with this method.
Therefore, we further improve on the cumulant calculation by measuring histograms in the polymer number $Q$ and directly extract moments $\mu_{2n}$ from it. In order to do so we define the trinomial distribution
\begin{align}
 D^{Q B}_{\mu_B/T}(Q,B)&=\sum_{P=|B|}^{Q}\binom{Q}{\frac{P+B}{2},\frac{P-B}{2},Q-P}
\frac{e^{B\mu_B/T}(\Nc+1)^{Q-P} }{w_Q(\mu_B/T)^Q},
\label{BinomTrinom}
\end{align}
which is used to get from histograms in the fully resummed polymer number $Q$ to histograms in the $B$
\begin{align}
 H^{B}_{V_\sigma,T,\mu_B}(B)&= \sum_{Q=P}^{V_\sigma}  D^{Q B}_{\mu_B/T}(Q,B)\, H^{Q}_{V_\sigma,T,\mu_B}(Q).
\end{align}
Finally, higher moments in the baryon number density are computed from the above histograms as 
\begin{align}
\langle f(B)\rangle =H^B_{V_\sigma,T,\mu_B}(B)\,f(B),
 \end{align}
 \begin{figure}[t]
\centering
\includegraphics[page=3,width=0.80\textwidth]{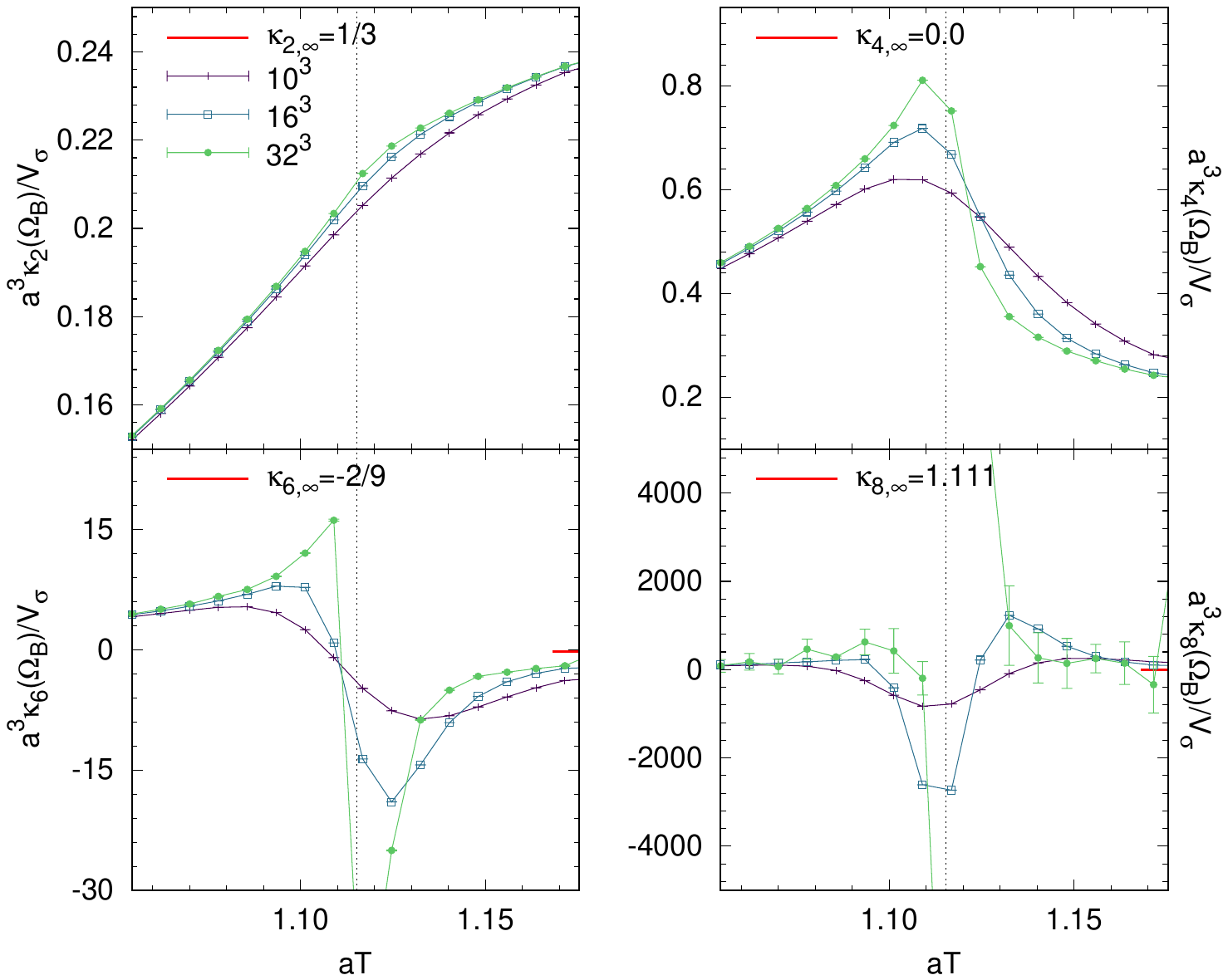}
\caption{Cumulants $\kappa_2$ to $\kappa_8$ for three different volumes in the vicinity of $T_c$. Finite volume effects become more dominant as the order in $\kappa$ is increased.}
\label{Fig:histogramk2TOk8}
\end{figure}
with $f(B)$ being the function that specifies which moment is measured. A great advantage of this method is that much higher moments can be extracted. In contrast, measuring baryon density moments directly with the CT algorithm necessarily needs the specification which moments are required. 

Applying the histogram method allows us to measure higher order cumulants even for lattices $N_\sigma^3\times CT$ with $N_\sigma=16$ and $N_\sigma=32$ (Fig.~\ref{Fig:histogramk2TOk8}). For simulations on $10^3\times CT$ lattices, cumulants up to $\kappa_{12}$ can be obtained with reasonable statistics. Nevertheless, as the volume is increased it becomes harder to keep the precision and $32^3\times CT$ simulations fail already at $\kappa_8$. It can be seen that the cumulants diverge at locations around $T_c$ when increasing the volume. 

\section{Discussion and Outlook}
We have presented a histogram method which vastly improves on precision in the calculation of cumulants. These are the main contribution to the Taylor coefficients defined in Eq.~(\ref{Eq:defineTayloCoeff}). The intrinsic feature of the CT limit of having static baryons results in a cleaner signal for the baryon density moments. Thus, a polymer resummation is natural. Besides the noise reduction we benefit from the histogram method by calculating arbitrary moments in $\Omega_B$ without performing new worm calculations.

Based on the cumulants a vast spectrum of analyses can be done. In \cite{Klegrewe2019} we presented results on the radius of convergence which tell us the location of the closest non-analyticity. Here, we are able to construct ratios $\kappa_2/\kappa_4, \dots, \kappa_{10}/\kappa_{12}$ and compare the resulting phase boundaries with the known phase diagram of SC-QCD. 
Furthermore, the pressure can be reconstructed at finite $\mu_B$ by a Taylor expansion of $\langle n_{D_s}\rangle$. This corresponds to the general - not necessarily homogeneous - definition of the pressure Eq.~(\ref{Eq:nativePressure}), which can be computed using methods as in \cite{Allton2005}.

All above calculations are performed in the chiral limit, so it would be interesting to introduce finite quark masses in the future. Furthermore, no $\beta$ corrections are introduced in CT simulations so far. However, these are necessary and can be incorporated in the same fashion as has been done for discrete time. Finally, we plan to go beyond one flavor calculations and implement a CT algorithm for $N_f=2$.

\section{Acknowledgments}
Numerical simulations were performed on the OCuLUS cluster at PC2 (Universität Paderborn). We acknowledge support by the Deutsche Forschungsgemeinschaft (DFG) through the Emmy Noether Program under Grant No. UN 370/1 and through the CRC-TR 211 'Strong-interaction matter under extreme conditions'– project number 315477589 – TRR 211.

\end{document}